\documentclass{Interspeech2024}
\usepackage{multirow}
\usepackage[table,xcdraw]{xcolor}
\usepackage{algorithm2e}
\usepackage{siunitx}
\usepackage{url}
\usepackage{hyperref}

\newcommand\blfootnote[1]{%
  \begingroup
  \renewcommand\thefootnote{}\footnote{#1}%
  \addtocounter{footnote}{-1}%
  \endgroup
}

\makeatletter
\newcommand{\removelatexerror}{\let\@latex@error\@gobble}
\makeatother

\RestyleAlgo{ruled}
\ninept

\interspeechcameraready

\title{Improving Generalization of Speech Separation in Real-World Scenarios: Strategies in Simulation, Optimization, and Evaluation}

\name[affiliation={1,2}]{Ke}{Chen}
\name[affiliation={2}]{Jiaqi}{Su}
\name[affiliation={1}]{Taylor}{Berg-Kirkpatrick}
\name[affiliation={1}]{Shlomo}{Dubnov}
\name[affiliation={2}]{Zeyu}{Jin}

\address{
  \vspace{-0.25cm}
  $^1$University of California San Diego, USA \quad
  $^2$Adobe Research, USA 
}
\email{\{knutchen, tberg, sdubnov\}@ucsd.edu, \{jsu, zejin\}@adobe.com}

\keywords{speech separation, data simulation, multi-loss optimization}

\begin{document}

\maketitle

\begin{abstract}
Achieving robust speech separation for overlapping speakers in various acoustic environments with noise and reverberation remains an open challenge.
Although existing datasets are available to train separators for
\textit{specific} scenarios, they do not effectively generalize 
across diverse real-world scenarios. In this paper, we present a novel data simulation pipeline that produces diverse training data from a range of acoustic environments and content, and propose new training paradigms to improve quality of a general speech separation model. Specifically, we first introduce AC-SIM, a data simulation pipeline that incorporates broad variations in both content and acoustics. Then we integrate multiple training objectives into the permutation invariant training (PIT) to enhance separation quality and generalization of the trained model. Finally, we conduct comprehensive objective and human listening experiments across separation architectures and benchmarks to validate our methods, demonstrating substantial improvement of generalization on both non-homologous and real-world test sets.
\blfootnote{*We thank the Institute for Research and Coordination in Acoustics and Music (IRCAM) and Project \texttt{REACH} for supporting this project.}

\end{abstract}

\vspace{-0.2cm}
\section{Introduction}

Speech source separation aims to extract the audio tracks of each individual speaker in an audio mixture. Deep neural networks have emerged as a modeling cornerstone for achieving high-quality separation results. State-of-the-art single-channel speech separation models typically employ an end-to-end architecture that directly processes the time-domain mixture samples, with prominent works such as ConvTasNet~\cite{luo2019convtasnet}, Dual-path RNN~\cite{luo2020dprnn}, DPTNet~\cite{jing2020dptnet}, WaveSplit~\cite{neli2021wavesplit}, Sepformer~\cite{cem2021sepformer}, and Mossformer~\cite{zhao2023mossformer}. Some models operating in the time-frequency domain, such as TF-GridNet~\cite{qiu2023tfdnet}, also demonstrated exceptional performance. The application scope of speech separation systems is vast, encompassing areas such as speech enhancement, speech translation, voice conversion, and beyond.


An inherent challenge in achieving robust speech separation lies in adapting the model to diverse real-world acoustic environments and conversational scenarios. Researchers have proposed datasets representing various environments, from clean acoustic environments (wsj0-2mix~\cite{hershey2016wsj02mix}), to noisy surroundings (WHAM!~\cite{mattew2020whamr}), and even the environment featuring both noise and reverberation (WHAMR!~\cite{mattew2020whamr} and SMS-WSJ~\cite{drude2019smswsj}). Furthermore, previous studies show drops in performance when separation models trained on datasets like wsj0-2mix are tested on comparable datasets. Multiple efforts have been made to address these issues, such as the introduction of the LibriMix~\cite{joris2020librimix} dataset that provides a wide range of speakers. Despite this promising progress, there remain open challenges in improving the robustness of speech separators.

\begin{figure*}[t]
    \centering
    \vspace{-0.5cm}
    \includegraphics[width=\textwidth]{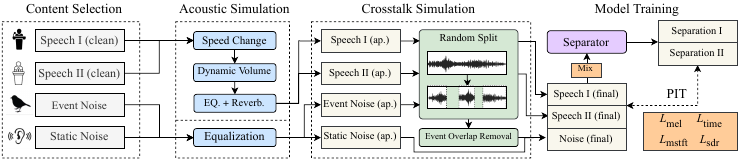}
    \vspace{-0.5cm}
    \caption{The complete pipeline of data simulation and training paradigm.}
    \label{fig:arch}
    \vspace{-0.3cm}
\end{figure*}

First, the generalization of speech separators across different \textit{acoustic environments} needs further exploration. 
Currently, researchers often train distinct models tailored to specific datasets such as wsj0-2mix, WHAM!, and WHAMR!. However, these datasets typically focus on singular acoustic environments, overlooking the complexity of real-world scenarios.
Hence, there is significant value and broad applicability in developing an all-in-one speech separator capable of handling any possible acoustic environment. Further, the remarkable advancements in speech enhancement and de-reverberation models~\cite{su2020hifigan, su2021hifigan2} provide an opportunity to reconsider the focus of speech separation in reverberant environments. Preserving the acoustics of the speech in the mixture (i.e., its reverberation and equalization) in separation results could offer substantial support for a variety of applications, including voice conversion and speech-to-speech translation. 

Second, the generalization of speech separators to diverse \textit{speech content} needs further exploration. Real-world speech scenarios span from dialogues where participants take turns speaking (e.g., interviews and presentations), to more volatile situations where speakers talk over each other (e.g., debates and talk shows). Each type of scenario presents unique characteristics, such as varying levels of cross-talk, presence of different non-speech audio events, levels of static noise, and instances of isolated single speakers. Developing a separator capable of generalizing across these variations in content requires careful consideration not only in data simulation techniques, but also in designing loss functions: traditional source-to-distortion ratio (SDR) and SI-SDR~\cite{roux2019sisdr} cannot explicitly optimize for single speaker samples or mixtures with little overlapping regions. Thus, it is essential to re-consider the loss function for achieving robust separation performance and desirable perceptual quality.


In this paper, we propose methods to address challenges: 

\begin{itemize}[leftmargin=*]
    \item We introduce \textbf{A}coustic-\textbf{C}ontent \textbf{Sim}ulation (AC-SIM), a comprehensive data simulation process that incorporates both acoustic and content variations during the training of speech separators. This approach aims to align separators to a diverse range of real-world speech separation scenarios.
    
    \item We integrate multiple training objectives into the permutation invariant training (PIT)~\cite{yu2017pit} scheme for speech separation to enhance the separation quality and model capabilities. 
    
    \item We conduct comprehensive objective and subjective experiments to demonstrate that the proposed methods improve the generalization of separators to multiple speech scenarios by achieving consistent performance on our evaluation sets, conventional benchmarks, and real-world cases.
\end{itemize}

\section{Methodology}

\subsection{Problem Formulation}

In this paper, we focus on single-channel speech source separation, addressing a common case involving up to two speakers. Given an audio mixture track $m \in [-1,1]^T$ with the sample length $T$, the model separates out a 2-channel track $\hat{s} \in [-1,1]^{2 \times T}$, with one channel for each speaker.

Prior work in speech separation defines the separation objective as isolating the \textbf{clean} speech tracks of individual speakers, typically under the assumption that noise is the primary interference to the speech mixture.
However, this separation objective becomes less effective in speech scenarios involving reverberation and equalization, as they introduce ambiguity in predicting clean speech and combine both separation and de-reverberation tasks into a single target.
%
%
Instead, we redefine the separation target across all scenarios as 
\textbf{speech tracks of individual speakers with acoustic characteristics} 
(i.e., preserving the reverberation and equalization correlated with speech, but suppressing noise). 
This target remains consistent with previous work under reverb-free clean or noisy scenarios but diverges in addressing noisy-reverberant scenarios. It clearly distinguishes the separation task from the enhancement task, thereby simplifying problem-solving approaches.
The preserved acoustic characteristics also benefit a wide range of downstream applications that require or can leverage their acoustic properties.

\subsection{Acoustic-Content Data Simulation}

We propose Acoustic-Content Simulation (AC-SIM) for data generation, comprising content and acoustic simulation processes. The content simulation contains two stages, namely content selection (section~\ref{sec:content-preselection}), and crosstalk simulation (section~\ref{sec:crosstalksim}). The acoustic simulation serves as the intermediate processing step to augment the selected content (section~\ref{sec:asim}).

\subsubsection{Content Selection} \label{sec:content-preselection}

Content selection is based on dynamic mixing (DM)~\cite{neli2021wavesplit}, which dynamically determines what samples to include in the audio mixture. We categorize three types of audio samples:
\begin{itemize}[leftmargin=*]
    \item Speech: clean speech samples of different speakers.
    \item Static Noise: generic noise samples as part of background.
    \item Event Noise: specific event samples as part of background.
\end{itemize}

In contrast to prior works~\cite{hershey2016wsj02mix,mattew2020whamr} which train models in one specific scenario, all samples in our selection, except the first speaker, are chosen with independent probabilities to maximize the diversity of speech scenarios. Furthermore, the inclusion of event noise provides a more dynamic sound background jointly with the static noise. This selection allows us to investigate whether a single separator can effectively handle multiple and complex speech scenarios.


\begin{figure}[t]
\removelatexerror
\begin{algorithm}[H]
\caption{Random Split}\label{alg:segmentsim}
\SetKwComment{Comment}{/* }{ */}
\SetKw{Constants}{Constants}
\SetKw{Input}{Input}
\SetKw{Output}{Output}
\SetKw{Variable}{Variable}
\Constants: $l_1=0.2$, $l_2=1.0$, $p_{seg}=0.75$ \\
\Input: audio samples $x \in [-1,1]^T$ \\
\Variable: audio output samples $y \gets \textbf{0} \in [-1,1]^T$ \\
$i \gets 0$, $j \gets 0$, $p \gets 0$ \\
\While{$p \le p_{seg}$ \& $i \le T$ \& $j \le T$}{
    $k \gets \text{randomint}(l_1 \times (T-i), l_2 \times (T-i))$ \\
    $j \gets \text{randomint}(j, T)$, $k \gets \text{min}(k, T-j)$ \\
    $y[j:j+k] \gets x[i:i+k]$, $i \gets i+k$, $j \gets j+k$ \\ 
    $p \gets \text{randomfloat}(0, 1)$
}
\Output: $y$
\end{algorithm}
\vspace{-0.5cm}
\end{figure}



\subsubsection{Acoustic Simulation} \label{sec:asim}
The acoustic simulation involves manipulating audio samples based on their types. The middle part of Figure~\ref{fig:arch} illustrates the use of speed change, dynamic volume normalization, reverberation and equalization. Similar to content selection, all acoustic manipulations are randomly chosen with probabilities to enrich the training data with diverse acoustic environments, thus enhancing the adaptability of the separator.

The speed change module, dynamic volume normalization and reverberation are applied exclusively to speech samples. The reverberation is guided by a randomly selected room impulse response (RIR) and augmented by scaling DRR and RT60 properties~\cite{bryan2019reverb} from 0.5 to 2.0. Speed change module stretches speech samples with a scale of 0.9 to 1.2. Dynamic volume normalization involves randomly inserting 0-3 anchor points and linearly changing the volume level to the target level in the range of -10 to 10 dB. 
Equalization (EQ) is applied to all speech, static noise, and event noise samples, employing a random seven-band filter with the gain for each band set between -5 and 5 dB. Notably, we applied an independent EQ before reverberation on speech to increase the simulation diversity.



\begin{table*}[t]
\centering
\vspace{-0.5cm}
\caption{The SI-SDRi / Silence-SDRi performance across 3 separation backbones on 12 evaluation sets. The trend of color from light to dark across DM, AC-SIM, and AC-SIM-ML demonstrates the continuous improvements.}
\label{tab:main_result}
\resizebox{0.95\textwidth}{!}{
\begin{tabular}{lcccccccccccc}
\toprule
 & \multicolumn{12}{c}{SI-SDRi / Silence-SDRi (dB $\uparrow$) on proposed evaluation sets and conventional benchmarks } \\ \cmidrule{2-13} 
\multirow{-2}{*}{Model} & D-All & D-NE & D-NR & D-N & S-All & S-NE & S-NR & \multicolumn{1}{c|}{S-N} & wsj0-2mix & WHAM! & WHAMR!* & Libri2Mix \\ \midrule
Sepformer - Vanilla & 5.80 & 12.23 & 6.01 & 12.73 & 5.66 & 14.03 & -1.02 & \multicolumn{1}{c|}{7.53} & \textbf{12.59} & \textbf{13.46} & \textbf{11.07} & 11.21 \\
Sepformer - DM & 10.73 & 16.13 & 11.23 & 17.13 & 23.89 & 27.49 & 19.13 & \multicolumn{1}{c|}{24.08} & \cellcolor[HTML]{ECF4FF}10.82 & \cellcolor[HTML]{ECF4FF}11.73 & \cellcolor[HTML]{ECF4FF}9.02 & \cellcolor[HTML]{ECF4FF}11.35 \\
Sepformer - AC-SIM & 13.05 & 18.78 & 13.73 & 20.05 & 22.06 & 29.12 & 19.01 & \multicolumn{1}{c|}{25.99} & \cellcolor[HTML]{DAE8FC}12.35 & \cellcolor[HTML]{DAE8FC}12.05 & \cellcolor[HTML]{DAE8FC}9.33 & \cellcolor[HTML]{DAE8FC}11.49 \\
Sepformer - AC-SIM-ML & \textbf{13.33} & \textbf{19.06} & \textbf{13.99} & \textbf{20.54} & \textbf{43.88} & \textbf{42.80} & \textbf{41.48} & \multicolumn{1}{c|}{\textbf{40.17}} & \cellcolor[HTML]{ADCCF8}12.36 & \cellcolor[HTML]{ADCCF8}12.26 & \cellcolor[HTML]{ADCCF8}9.48 & \cellcolor[HTML]{ADCCF8}\textbf{11.74} \\ \midrule
ConvTasNet - Vanilla & 2.87 & 8.89 & 2.89 & 9.53 & 3.46 & 8.45 & -2.76 & \multicolumn{1}{c|}{2.57} & \textbf{9.78} & \textbf{10.72} & \textbf{8.41} & 8.30 \\
ConvTasNet - DM & 6.40 & 12.19 & 6.90 & 12.86 & 20.60 & 21.92 & 15.86 & \multicolumn{1}{c|}{18.21} & \cellcolor[HTML]{ECF4FF}6.53 & \cellcolor[HTML]{ECF4FF}6.72 & \cellcolor[HTML]{ECF4FF}4.93 & \cellcolor[HTML]{ECF4FF}6.95 \\
ConvTasNet - AC-SIM & 9.94 & 15.32 & 10.48 & 16.44 & 10.40 & 13.92 & 5.80 & \multicolumn{1}{c|}{9.96} & \cellcolor[HTML]{DAE8FC}8.37 & \cellcolor[HTML]{DAE8FC}8.00 & \cellcolor[HTML]{DAE8FC}5.88 & \cellcolor[HTML]{DAE8FC}8.46 \\
ConvTasNet - AC-SIM-ML & \textbf{10.86} & \textbf{16.23} & \textbf{11.40} & \textbf{17.47} & \textbf{34.54} & \textbf{32.87} & \textbf{32.10} & \multicolumn{1}{c|}{\textbf{29.43}} & \cellcolor[HTML]{ADCCF8}8.99 & \cellcolor[HTML]{ADCCF8}8.49 & \cellcolor[HTML]{ADCCF8}6.32 & \cellcolor[HTML]{ADCCF8}\textbf{8.95} \\ \midrule
DPRNN - Vanilla & 1.92 & 9.79 & 2.18 & 10.79 & 5.60 & 11.62 & -0.92 & \multicolumn{1}{c|}{7.20} & \textbf{10.06} & \textbf{11.58} & \textbf{9.30} & 9.14 \\
DPRNN - DM & 7.09 & 11.82 & 7.40 & 12.65 & 8.23 & 11.77 & 2.65 & \multicolumn{1}{c|}{7.00} & \cellcolor[HTML]{ECF4FF}7.35 & \cellcolor[HTML]{ECF4FF}7.19 & \cellcolor[HTML]{ECF4FF}5.21 & \cellcolor[HTML]{ECF4FF}7.61 \\
DPRNN - AC-SIM & 9.59 & 15.45 & 9.95 & 16.72 & 27.86 & 28.05 & 24.78 & \multicolumn{1}{c|}{25.79} & \cellcolor[HTML]{DAE8FC}9.61 & \cellcolor[HTML]{DAE8FC}8.08 & \cellcolor[HTML]{DAE8FC}5.86 & \cellcolor[HTML]{DAE8FC}9.22 \\
DPRNN - AC-SIM-ML & \textbf{10.88} & \textbf{16.36} & \textbf{11.59} & \textbf{17.86} & \textbf{43.21} & \textbf{41.39} & \textbf{42.62} & \multicolumn{1}{c|}{\textbf{40.62}} & \cellcolor[HTML]{ADCCF8}9.94 & \cellcolor[HTML]{ADCCF8}8.99 & \cellcolor[HTML]{ADCCF8}6.69 & \cellcolor[HTML]{ADCCF8}\textbf{9.73} \\ \bottomrule
\end{tabular}}
\vspace{-0.3cm}
\end{table*}

\subsubsection{Crosstalk Simulation} \label{sec:crosstalksim}

The crosstalk simulation, as part of content simulation, is applied on acoustically-processed audio samples (as \textbf{ap.}). It consists of two steps: random split and event overlap removal. 

Most speech datasets contain continual speech without long pauses between utterances, whereas real-world conversations often involve turn-taking (i.e., crosstalk) and silent moments when speakers are not actively engaged. Naively mixing speech samples from these datasets results in excessive overlap of speakers, creating a distribution different from real-world data.

Instead, we propose randomly splicing segments of speech together with intervals of silence, as shown in Algorithm~\ref{alg:segmentsim}: for each audio component $x$, we randomly draw a segment with a random length between $l_1 \times T$ and $l_2 \times T$ from $x$ and append it to the output samples $y$. The process continues each time with probability $p_{seg}$ until either input or output samples reach their end. In Algorithm~\ref{alg:segmentsim}, we also describe how splitting and splicing operations are randomly conducted to create crosstalk scenes.

Another content factor we consider is whether speech and non-speech events are disjoint. Some transitions occur in scenarios like talk shows or interviews where speakers remain silent until non-speech events cease. Therefore, when mixing segmented speech samples with event samples, we implement an event overlap removal mechanism with a certain probability. This involves removing all the segments in the event track that overlap with speech occurrences.
By incorporating crosstalk simulation, our training data more accurately reflects the dynamics between speech and non-speech events in real-world scenarios, enhancing the robustness of the separator.

\begin{table}[t]
\caption{The SI-SDRi of Sepformer with different ablations.}
\label{tab:aba_result}
\resizebox{\columnwidth}{!}{
\begin{tabular}{lccccc}
\toprule
\multirow{2}{*}{Sepformer} & \multicolumn{5}{c}{SI-SDRi (dB $\uparrow$)} \\ \cmidrule{2-6} 
 & D-All & D-NE & D-NR & D-N & Libri2Mix \\ \midrule
w/. None & 1.81 & 7.90 & 2.04 & 9.88 & 4.64  \\
w/. A-SIM & 7.14  & 12.30 & 7.13 & 14.77 & 10.48  \\
w/. C-SIM & 1.78 & 17.61 & 1.34 & 18.90 & 10.50  \\
w/. AC-SIM & 13.05 & 18.78 & 13.73 & 20.05 & 11.49 \\
w/. AC-SIM-ML & 13.33 & 19.06 & 13.99 & 20.54 & 11.74 \\ \bottomrule
\end{tabular}}
\vspace{-0.5cm}
\end{table}

\begin{table*}[t]
\centering
\vspace{-0.5cm}
\caption{The performance of subjective listening test, blue scores represent the top models in each dataset without significant difference.}
\label{tab:mos_result}
\resizebox{0.95\textwidth}{!}{
\begin{tabular}{lcccccccc}
\toprule
 & \multicolumn{8}{c}{Mean Opinion Score} \\ \cmidrule{2-9} 
\multirow{-2}{*}{Model} & D-All & S-All & wsj0-2mix & WHAMR! & Libri2Mix-Clean & Libri2Mix-Noise & \multicolumn{1}{c|}{Real Cases} & Average \\ \midrule
Sepformer - Vanilla & 1.84 $\pm$ 0.05 & 2.05 $\pm$ 0.06 & 3.87 $\pm$ 0.06 & \cellcolor[HTML]{ADCCF8}3.28 $\pm$ 0.06 & 3.49 $\pm$ 0.06 & \cellcolor[HTML]{ADCCF8}3.22 $\pm$ 0.06 & \multicolumn{1}{c|}{1.96 $\pm$ 0.07} & 2.85 $\pm$ 0.03 \\
Sepformer - DM & 2.56 $\pm$ 0.07 & 2.75 $\pm$ 0.07 & 3.94 $\pm$ 0.05 & 2.75 $\pm$ 0.06 & 3.90 $\pm$ 0.05 & 3.04 $\pm$ 0.06 & \multicolumn{1}{c|}{2.53 $\pm$ 0.07} & 3.08 $\pm$ 0.03 \\
Sepformer - AC-SIM & 2.88 $\pm$ 0.06 & 2.94 $\pm$ 0.08 & \cellcolor[HTML]{ADCCF8}3.97 $\pm$ 0.05 & 2.90 $\pm$ 0.06 & \cellcolor[HTML]{ADCCF8}4.08 $\pm$ 0.05 & 3.08 $\pm$ 0.06 & \multicolumn{1}{c|}{2.75 $\pm$ 0.08} & 3.24 $\pm$ 0.03 \\
Sepformer - AC-SIM-ML & \cellcolor[HTML]{ADCCF8}3.03 $\pm$ 0.06 & \cellcolor[HTML]{ADCCF8}3.38 $\pm$ 0.08 & \cellcolor[HTML]{ADCCF8}4.05 $\pm$ 0.05 & 2.88 $\pm$ 0.06 & \cellcolor[HTML]{ADCCF8}4.05 $\pm$ 0.05 & \cellcolor[HTML]{ADCCF8}3.27 $\pm$ 0.06 & \multicolumn{1}{c|}{\cellcolor[HTML]{ADCCF8}3.03 $\pm$ 0.08} & \cellcolor[HTML]{ADCCF8}3.39 $\pm$ 0.03 \\ \midrule
Mixture & 1.51 $\pm$ 0.04 & 2.18 $\pm$ 0.06 & 1.61 $\pm$ 0.05 & 1.50 $\pm$ 0.05 & 1.64 $\pm$ 0.06 & 1.43 $\pm$ 0.04 & \multicolumn{1}{c|}{2.20 $\pm$ 0.09} & 1.71 $\pm$ 0.02 \\
GroundTruth & 3.97 $\pm$ 0.06 & 3.84 $\pm$ 0.08 & 4.35 $\pm$ 0.04 & 4.27 $\pm$ 0.05 & 4.48 $\pm$ 0.04 & 4.64 $\pm$ 0.03 & \multicolumn{1}{c|}{---} & 4.03 $\pm$ 0.03 \\ \bottomrule
\end{tabular}}
\end{table*}
\begin{figure*}[t]
    \centering
    \vspace{-0.2cm}
    \includegraphics[width=\textwidth]{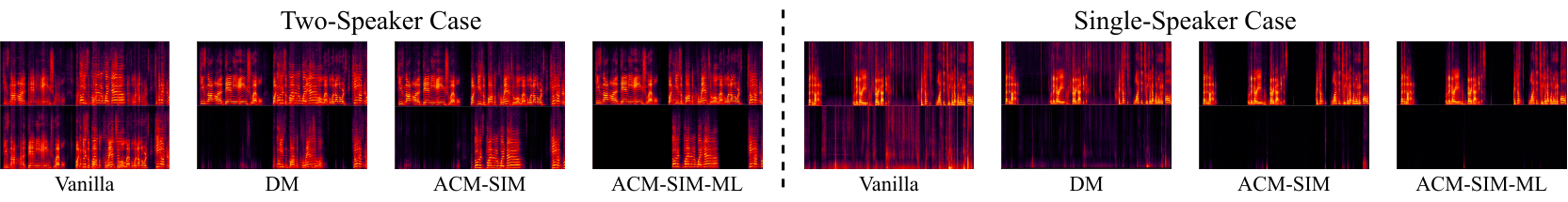}
    \caption{The visualization of spectrograms on separation results across different Sepformer models.}
    \label{fig:visual}
    \vspace{-0.5cm}
\end{figure*}


\subsection{Training Objectives and PIT}
Speech separation models are typically optimized with SDR or SI-SDR loss functions. However, neither explicitly optimizes for silence frames, which frequently occur in single-speaker or sparse talking scenarios, nor prioritizes the clarity of separation results. 
Inspired by \cite{kong2020hifigan,su2020hifigan,kim2021mdxnet}, we introduce three additional loss functions that better align with human perception of acoustic similarity into Permutation Invariant Training (PIT): multi-resolution STFT magnitude loss, mel-spectrogram loss, and time-domain L2 loss:

\begin{footnotesize}
\vspace{-0.3cm}
\begin{align}
    L_{\text{mstft}}&= \sum\nolimits_{i} \lVert{ 
    \text{ log } \vert \text{STFT}(s; \theta_i) \vert - \text{log } \vert  \text{STFT} (\hat{s}; \theta_i) \vert \rVert}_1  \\
    L_{\text{mel}}&= \lVert{ 
    \text{ log } \text{Mel}(s)  - \text{log }  \text{Mel} (\hat{s}) \rVert}_1 \\
    L_{\text{time}}&= \lVert{s - \hat{s}\rVert}_2 
\end{align}
\end{footnotesize}%
We adopt the multi-resolution STFT magnitude loss with three filter lengths: 512, 1024, and 2048. The mel-spectrogram loss is conducted using the STFT with a filter length 1024 and 128 mel-bins. The total loss for training the separator is the combination: $L= \lambda_{\text{mstft}} L_{\text{time}} + \lambda_{\text{mstft}} L_{\text{mstft}} + \lambda_{\text{mel}} L_{\text{mel}} +  \lambda_{\text{sdr}} L_{\text{sdr}}$. We use coefficients $\lambda_{\text{mstft}}$ = 10, $\lambda_{\text{mel}}$ = 10, $\lambda_{\text{time}}$ = 100, and $\lambda_{\text{sdr}}$ = 1. We replace $L_{sdr}$ with the combined loss $L$ during the permutation invariant training (PIT) of the speech separator. 

\section{Experiments}

\subsection{Datasets, Backbones, and Training Setup}
We train and validate speech separators on simulated samples at a sample rate of \SI{16000}{\hertz}. We collect 6149 hours speech data from VCTK~\cite{veaux2013vctk}, DAPS~\cite{mysore2015daps} and Spotify Podcast datasets~\cite{yun2024gr0}\footnote{6127 hours cleaned and filtered speech from podcast episodes.}; 4.36 hours static noise data from the REVERB Challenge~\cite{kinoshita2013reverb}, the ACE Challenge~\cite{eaton2016estimation} and the ambient sound databases~\cite{chen2021structure}; 39 hours ten-class event noise data\footnote{alarm, applause, bird, cough, cry, engine, laugh, pet, traffic, typing.} 
from ESC-50~\cite{piczak2015esc50}, BBC Sound Effects~\cite{laionclap2023}, VGGSound~\cite{chen2020vggsound}, UrbanSound~\cite{salamon2014us8k}, and Freesound datasets~\cite{font2013freesound}; and RIR data for reverberation from MIT IR Survey dataset~\cite{traer2016mitir} and the Echothief IR Library~\cite{EchoThief_2021}. Each of them is split into the training set and the validation/evaluation sets. 

To fully validate the effectiveness of AC-SIM and training paradigm, we choose three models from the prominent and easily-accessible speech separation work as our backbones: ConvTasNet, Dual-Path RNN, and Sepformer. They represent common choices of network architectures for the speech separation task. 
We then train these models using several variants of our proposed data simulation.
Each training data sample lasts 5.0 seconds. We use the Adam~\cite{kingma2014adam} optimizer ($\beta_1$=0.9, $\beta_2$=0.99) for training all separators with the batch size of 16 and up to \num{200000} steps until convergence on 4 NVIDIA A100 GPUs. Due to the page limitation, we attach the hyper-parameters of these models, detailed configurations of AC-SIM and separation demos in the anonymous appendix page\footnote{\href{https://speechsep-acsim.github.io/}{https://speechsep-acsim.github.io/}}. 

To fully evaluate the separation performance of the models in different scenarios, we use AC-SIM to generate validation and evaluation sets shown in the leftmost part of Table~\ref{tab:main_result}, with the total duration of 6.8 hours. The prefixes \textbf{D} and \textbf{S} denote 2-speaker (double) and 1-speaker samples (single) respectively, and the suffixes denote different acoustic scenarios: \textbf{All} (static noise, event noise, and reverberant speech); \textbf{NE} (static noise, event noise, and clean speech); \textbf{NR} (static noise and reverberant speech); and \textbf{N} (static noise and clean speech). Necessarily, we assess the generalization of the models using wsj0-2mix, WHAM!, WHAMR!, and Libri2Mix (both clean and noisy versions) evaluation sets as benchmarks.

\subsection{Objective Evaluations and Ablation Studies}

Table~\ref{tab:main_result} presents the primary evaluation of three separation models across 12 evaluation sets. The ``Vanilla" models serve as a baseline and are trained and validated on the WHAMR! dataset with dynamic mixing. The ``DM" Model are trained with the same setting as WHAMR! (noisy and reverberant speech) but using our data sources. The ``AC-SIM" and ``AC-SIM-ML" models are trained with our simulation pipeline and data sources, with ``ML" incorporating the multiple loss objectives.
We employ the SI-SDR improvement metric to evaluate the separation performance. Notably, single-speaker cases cannot be evaluated using SI-SDR, we refer to use Silence-SDR improvement~\cite{kong2020silencesdr} to measure the separation quality in the speaker channel and silence in another channel:

\begin{footnotesize}
\vspace{-0.3cm}
\begin{align}
    \text{Silence-SDR}(s, \hat{s})=10\log\frac{||s||^2}{||\hat{s}||^2}    
\end{align}   
\end{footnotesize}

Comparing between Vanilla and DM models, we observe that Vanilla models perform well on wsj0-12mix, WHAM! and WHAMR!, which share the same speech data sources. However, they exhibit decreased performance on our evaluation sets
Similarly, the DM models perform well on our evaluation sets but struggle to generalize to conventional benchmarks. This demonstrates that simply changing the data source, even increasing the data size, without thorough simulation does not significantly improve separation performance, leading to challenges in generalization across non-homologous evaluation sets.

Tracking performance among DM, AC-SIM, and AC-SIM-ML models, we observe a consistent improvement trend across wsj0-2mix, WHAM!, WHAMR!, and Libri2Mix, denoted by the blue color progression. This highlights the effectiveness of our proposed AC-SIM, enhancing the generalization ability of models to overcome limitations of training sources.
Furthermore, employing multiple loss objectives (ML) during training leads to substantial improvements, especially on \textbf{S} sets, enhancing the capability in single-speaker cases and overall separation quality.
Ultimately, our AC-SIM-ML models demonstrate superior performance on our evaluation sets across different cases, gradually approach the performance of Vanilla models on conventional benchmarks, and achieve the best performance on Libri2Mix, an external evaluation set that none of the four models encountered during training. This underscores the robust generalization offered by AC-SIM and multi-loss training.

We also present the ablation study results in Table~\ref{tab:aba_result}, by training the Sepformer model with acoustic-only simulation (\textbf{A-SIM}), content-only simulation (\textbf{C-SIM}), AC-SIM, multiple objectives, and only clean speech mixtures (\textbf{None}). We observe clear differences on the SI-SDRi performance across evaluation sets. The \textbf{A-SIM} Sepformer demonstrates strong performance on D-NR and D-All sets, excelling in handling reverberant speech. However, it under-performs compared to the \textbf{C-SIM} Sepformer when faced with static noise and event noise (D-N and D-NE). Conversely, the \textbf{C-SIM} Sepformer struggles with reverberant cases and shows no improvement over the \textbf{None} Sepformer D-NR and D-All sets. Combining all simulations and applying multiple objectives consistently improves model performance across all speech separation cases.

\vspace{-0.2cm}
\subsection{Subjective Listening Test}

We evaluate the separation accuracy and perceptual quality with a Mean Opinion Score (MOS) test on the evaluation sets as well as a collection of 16 real-world speech recordings\footnote{The real-world recordings are attached in our demo website.}, and present results in Table~\ref{tab:mos_result}. The test was conducted on 
Prolific~\cite{Proflic}. In each question, a listener rates the separation quality of two isolated tracks individually on a scale from 1 (Bad) to 5 (Excellent) with the original mixture track provided as a reference. 
We collected MOS scores of the four Sepformer models in Table~\ref{tab:main_result}, along with the input mixture and the ground-truth reference, totalling 13401 valid responses from 275 unique listeners ($\sim$2200 responses per model).

From the table, our ACM-SIM-ML Sepformer outperforms other models across all evaluation sets and real-world cases except WHAMR!, with a consistent improvement from DM to ACM-SIM-ML setting. Conversely, the Vanilla Sepformer, trained on WHAMR!, only fits its homologous evaluation set but performs poorly on the others. This indicates: (1) WHAMR! may not be a representative dataset for real-world scenarios, as models excelling in real-world cases and other evaluation sets tend to under-perform on WHAMR!, and vice versa; (2) Our AC-SIM helps generalize the model to real-world cases as well as different benchmarks via a comprehensive simulation process. 
Besides, the visualizations of a real-world example in Figure~\ref{fig:visual} further demonstrates that AC-SIM-ML enhances clarity of audio and rectifies issues encountered in single-speaker cases. 

\vspace{-0.3cm}
\section{Conclusion} 
We propose a combination of acoustic-content simulation, and multiple loss training paradigms to address the challenge of generalizing speech separation models across diverse acoustic environments and content. Our experiments confirm the effectiveness of our methods across various models and benchmarks, showcasing substantial improvements in separation quality and generalization on both synthetic and real-world test sets. This underscores further research to bridge the gap between speech separation technologies and real-world applications.


\bibliographystyle{IEEEtran}
\bibliography{mybib}

\end{document}